\documentclass[compactaffiliation]{Interspeech}

\interspeechcameraready

\title{Accelerating Flow-Matching-Based Text-to-Speech \\
 via Empirically Pruned Step Sampling}

\author[affiliation={1}]{Qixi}{Zheng}
\author[affiliation={1,2}]{Yushen}{Chen}
\author[affiliation={1,2}]{Zhikang}{Niu}
\author[affiliation={1}]{Ziyang}{Ma}
\author[affiliation={3}]{Xiaofei}{Wang}
\author[affiliation={1}]{Kai}{Yu}
\author[affiliation={1,2,\dagger}]{Xie}{Chen}
\affiliation{MoE Key Lab of Artificial Intelligence, X-LANCE Lab, School of Computer Science}{\\Shanghai Jiao Tong University}{China}
\affiliation{}{Shanghai Innovation Institute}{China}
\affiliation{}{Microsoft}{USA}
\email{\{jerrister.150329, chenxie95\}@sjtu.edu.cn}

\usepackage{comment}

\usepackage{graphicx}
\usepackage{amsmath}
\usepackage{amssymb}
\usepackage{multirow}
\usepackage{booktabs}
\usepackage{subcaption}
\usepackage{arydshln}
\usepackage{stfloats}
\usepackage{color, xcolor}
\usepackage{hyperref}

\newcommand\blfootnote[1]{
  \begingroup
  \renewcommand\thefootnote{}\footnote{#1}
  \addtocounter{footnote}{-1}
  \endgroup
}

\begin{document}

\maketitle
\begin{abstract}
    % 1000 characters. ASCII characters only. No citations.
    Flow-matching-based text-to-speech (TTS) models, such as Voicebox, E2 TTS, and F5-TTS, have attracted significant attention in recent years. These models require multiple sampling steps to reconstruct speech from noise, making inference speed a key challenge. Reducing the number of sampling steps can greatly improve inference efficiency. To this end, we introduce Fast F5-TTS, a training-free approach to accelerate the inference of flow-matching-based TTS models. By inspecting the sampling trajectory of F5-TTS, we identify redundant steps and propose Empirically Pruned Step Sampling (EPSS), a non-uniform time-step sampling strategy that effectively reduces the number of sampling steps. Our approach achieves a 7-step generation with an inference RTF of 0.030 on an NVIDIA RTX 3090 GPU, making it 4 times faster than the original F5-TTS while maintaining comparable performance. Furthermore, EPSS performs well on E2 TTS models, demonstrating its strong generalization ability.

\end{abstract}

\blfootnote{$\dagger$ Corresponding author. }

\section{Introduction}

Recently, text-to-speech (TTS) has emerged as a promising approach for generating speech samples from a given input text 
while mimicking the voice characteristics of a reference speech.
Existing methods have achieved significant breakthroughs by scaling data and model size, enabling these systems to generate speech of high fidelity and naturalness that is nearly indistinguishable from real human voices. These approaches can be categorized into two types: auto-regressive models (AR)~\cite{Wang2023NeuralCL,lajszczak2024base,peng-etal-2024-voicecraft,anastassiou2024seed,meng2024autoregressive,du2024cosyvoice} and non-auto-regressive (NAR) models~\cite{le2024voicebox,Ju24naturalspeech,eskimez2024e2,wang2024maskgct,chen2024f5}.

Despite achieving high synthesis quality, the inference speed of current TTS models remains a challenge. 
AR models suffer from long inference latencies due to their sequential generation process. 
For instance, VALL-E~\cite{Wang2023NeuralCL} needs to sequentially predict 75 tokens when generating one second of speech.
Diffusion models fit well within the NAR framework for speech synthesis, which generates all tokens or the entire Mel spectrogram in a parallel fashion. However, multiple sampling steps are required to recover speech from noise by design. 
For example, NaturalSpeech 2~\cite{shen2023naturalspeech} takes 150 steps. 
In contrast,  flow matching with optimal transport (FM-OT)~\cite{lipman2022flow} improves efficiency by leveraging a direct transport trajectory, reducing the number of required steps.
Voicebox~\cite{le2024voicebox} introduced FM-OT into TTS systems with 64-step generation, and subsequent models like E2 TTS~\cite{eskimez2024e2} and F5-TTS~\cite{chen2024f5} further refined the approach, achieving high-quality synthesis in 32 steps.
There are consistent research efforts to further improve the inference costs of diffusion-based TTS models. 
NaturalSpeech 3~\cite{Ju24naturalspeech} lowers the number of steps to 30 using a factorized diffusion model.
FlashSpeech~\cite{ye2024flashspeech} achieves a 2-step generation through the latent consistency model (LCM)~\cite{luo2023latent}. Additionally, 
Distillation methods, such as rectified flow~\cite{liu2022flow} and distribution matching distillation (DMD)~\cite{yin2024one, yin2024improved}, have also proven effective in accelerating inference for pre-trained TTS models~\cite{guo2024voiceflow, liu2024autoregressive, li2024dmdspeech}.

In this paper, we mainly focus on the flow matching based TTS model, such as E2 TTS and F5-TTS, given their simplicity and promising performance. We propose Fast F5-TTS, a training-free, plug-and-play approach that accelerates inference in FM-OT-based TTS models through a novel Empirically Pruned Step Sampling (EPSS) strategy. 
EPSS is a non-uniform time-step sampling strategy that prunes unnecessary steps, retaining only those that are empirically shown to be essential. 
While sway sampling (SS)~\cite{chen2024f5} in F5-TTS enhances performance at a fixed number of function evaluations (NFE), EPSS focuses on maintaining model performance while further reducing the NFE. 
These two sampling strategies are shown to be complementary and can be effectively combined. In addition, we extend EPSS to the E2 TTS model, demonstrating its generalization ability and potential applicability to other flow-matching-based generative models. An online demo can be found at \url{https://fast-f5-tts.github.io/}. We will release our updated code incorporating the improvements in this paper on the existing F5-TTS GitHub repository \url{https://github.com/SWivid/F5-TTS}.

\section{Preliminaries}

\subsection{Formulation of flow matching}

Let $\mathbb{R}^d$ be the data space and $q$ be the real data distribution. 
Flow matching learns a time-dependent vector field $v_t: [0, 1] \times \mathbb{R}^d \rightarrow \mathbb{R}^d$, which generates a flow $\phi_t$ that transforms samples from a simple prior distribution $p_0$ to the target data distribution $q$ via $p_1 \approx q$.
This transformation is achieved by solving the ODE in Equation~\ref{eq: ODE} over the interval $t \in [0, 1]$:
\begin{equation}
    \label{eq: ODE}
    \text{d} \phi_t(x) = v_t(\phi_t(x)) \text{d} t, \quad \phi_0(x) = x.
\end{equation}
Let $p_t: [0, 1] \times \mathbb{R}^d \rightarrow \mathbb{R}^{>0}$ be the probability path of $\phi_t$.
The vector field $v_t$ is parameterized by a neural network $\theta$ and is regressed by the conditional flow matching (CFM) Loss
\begin{equation}
    \label{eq: cfmloss}
    \mathcal{L}_{\text{CFM}}(\theta) = \mathbb{E} \left \Vert v_t(x; \theta) - u_t(x | x_1) \right \Vert ^2,
\end{equation}
where $t \sim \mathcal{U}[0, 1]$, $x_1 \sim q(x_1)$, $ x \sim p_t(x | x_1)$, and $u_t$ is the vector field that transforms $p_0$ into $q$.

\subsection{Flow-matching-based TTS}

Flow-matching-based TTS can be divided into two main approaches: Mel spectrogram generation and speech token generation. As F5-TTS is employed as our baseline model, we mainly focus on the Mel spectrogram generation approach in this paper.

\noindent \textbf{Modeling and training.} 
The model is trained on the text-guided speech-infilling task~\cite{le2024voicebox}. 
The infilling task is to predict a segment of speech $m \odot x_1$ given its surrounding speech $(1 - m) \odot x_1$ and full text $z$ as the condition $c$. 
Using the optimal transport path $\phi_t(x) = (1 - t) x + t x_1$, we rewrite the CFM Loss as
\begin{equation}
    \label{eq: cfmloss2}
    \mathcal{L}_{\text{CFM}}(\theta) = \mathbb{E} \left \Vert m \odot (v_t(x_t, c; \theta) - (x_1 - x_0)) \right \Vert ^2,
\end{equation}
where $m$ is a mask and $x_t = (1 - t) x_0 + t x_1$.

\noindent \textbf{Inference.}
Given a sampled noise $x_0$ and conditions $c$ consisting of the text to generate and the audio prompt, the target $x_1 = \phi_1(x_0)$ can be calculated by solving the ODE in Equation~\ref{eq: ODE}:
\begin{equation}
    \label{eq: Solve ODE}
    x_1 =  x_0 + \int_0^1 v_t(\phi_t(x_0), c; \theta) \text{d} t.
\end{equation}
ODE solvers, such as the Euler method or the midpoint method, are employed to numerically solve this ODE. 
The number of function evaluations (NFE) refers to how many times the solver evaluates the neutral network $v_t(\cdot; \theta)$. As classifier-free-guidance (CFG)~\cite{ho2022classifier} is universally applied, we do not report NFE as doubled despite the use of CFG.
These solvers divide the continuous time interval $t \in [0,1]$ into a sequence of steps $\{t_k\}_{k=0}^{\text{NFE}}$, where $t_0 = 0$, $t_\text{NFE} = 1$.
The choice of time-steps $\{t_k\}$ is determined by a time-step scheduling policy $\pi(t)$, which defaults to a uniform schedule. 

\subsection{F5-TTS}

F5-TTS is a fully NAR TTS system based on FM-OT with diffusion transformer (DiT)~\cite{peebles2023scalable}. It employs a literally text-in-speech-out pipeline and adopts ConvNeXt~\cite{woo2023convnext} for text modeling, achieving state-of-the-art performance in zero-shot TTS.

F5-TTS introduced sway sampling, a non-uniform schedule which improves sampling performance, achieving higher efficiency at the same NFE. The sway sampling function 
\begin{equation}
    \label{eq:ss}
    \text{SS}(t;s) = t + s \cdot (\cos{(\frac{\pi}{2}t)} + t - 1)
\end{equation}
sways the time schedule to the left with coefficient $s \in [-1, 0)$, or to the right with $s \in (0, \frac{2}{\pi - 2}]$.

\begin{figure}[t]
  \centering
  \includegraphics[width=0.83\linewidth]{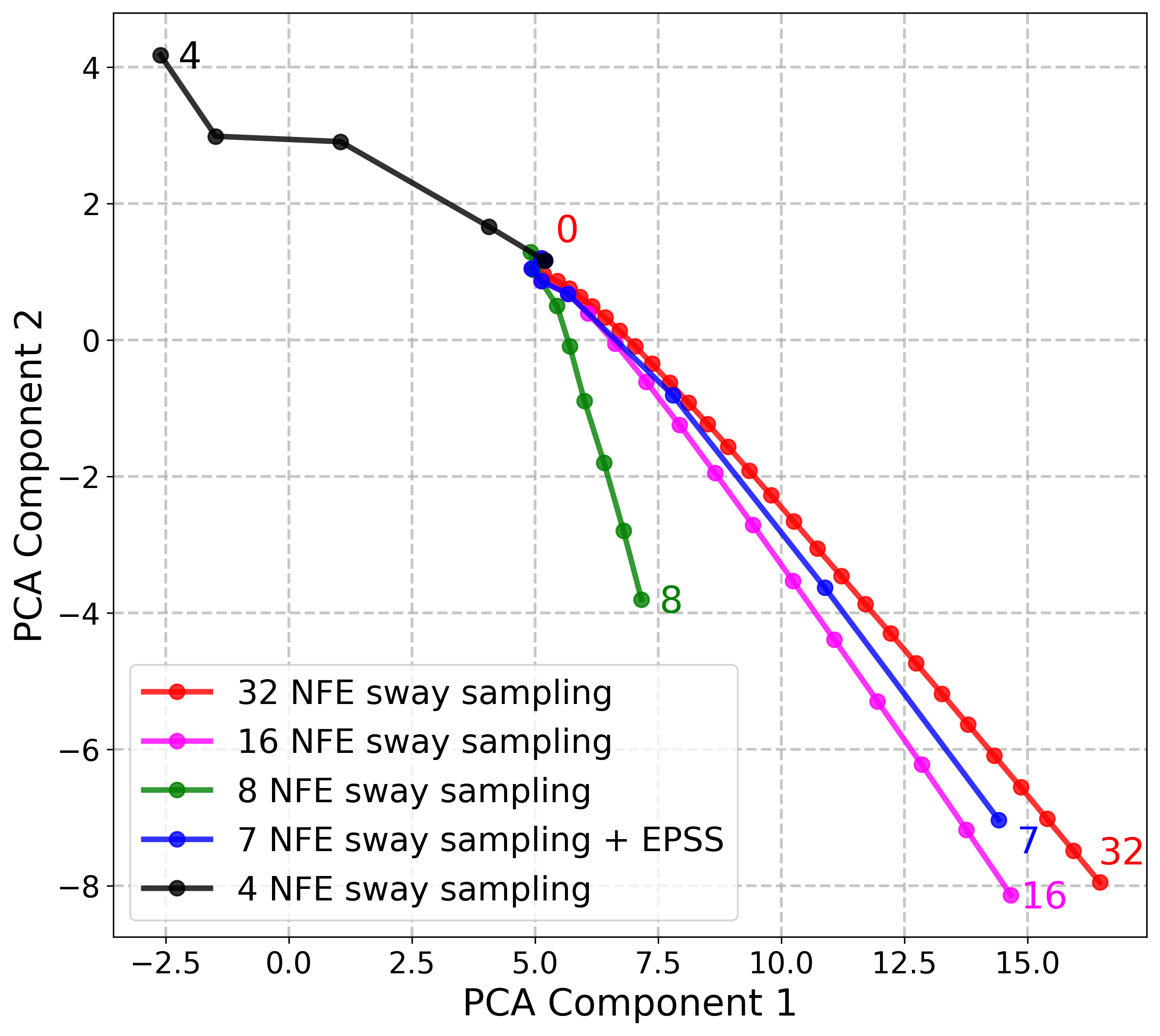}
  \caption {PCA visualization of the 100-dimensional sampling trajectory during the inference process of Mel spectrogram. 
  }
  \label{fig:traj_base}
\end{figure}

\begin{figure*}[t]
  \centering
  \includegraphics[width=0.32\linewidth]{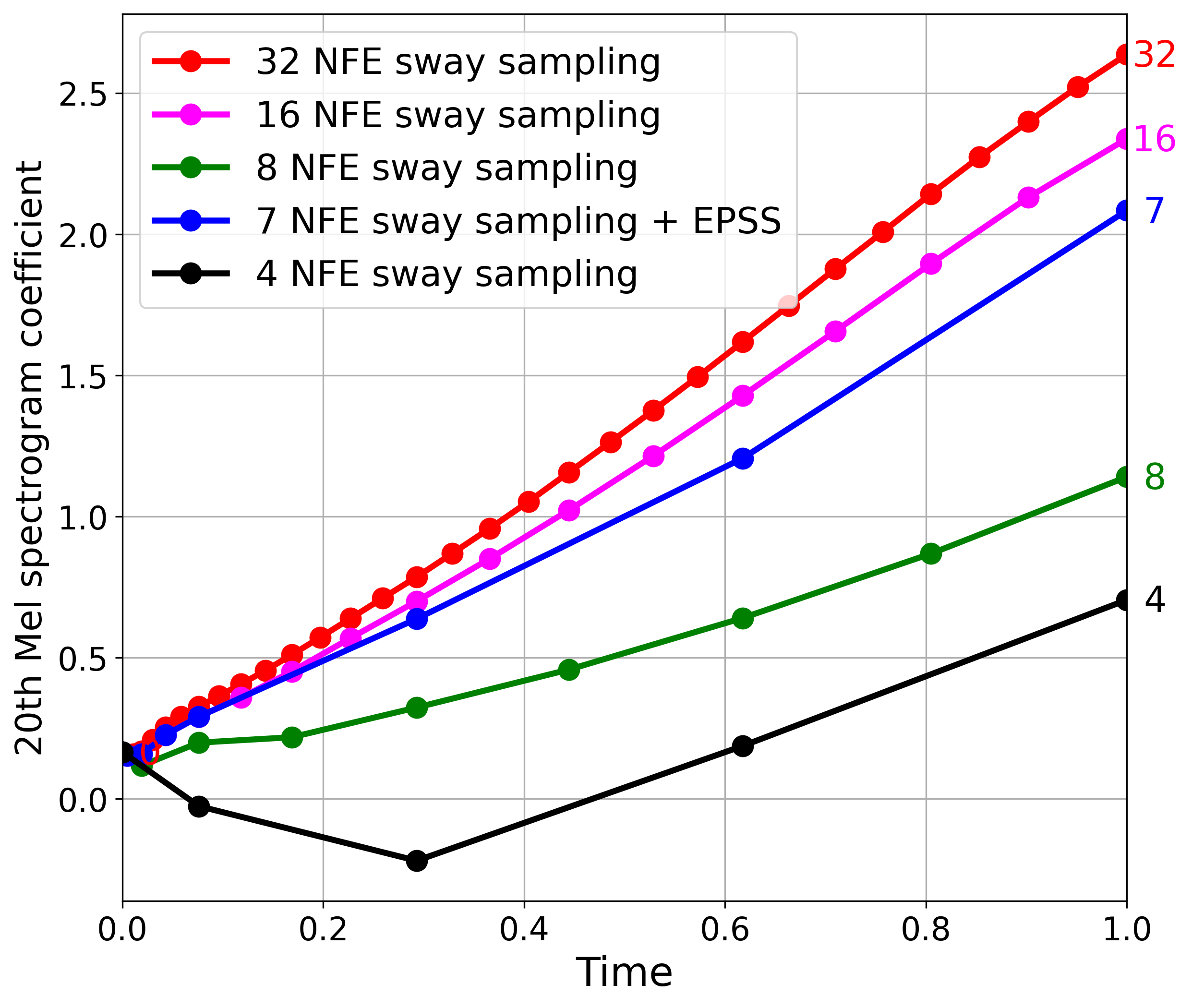} \hfill
  \includegraphics[width=0.32\linewidth]{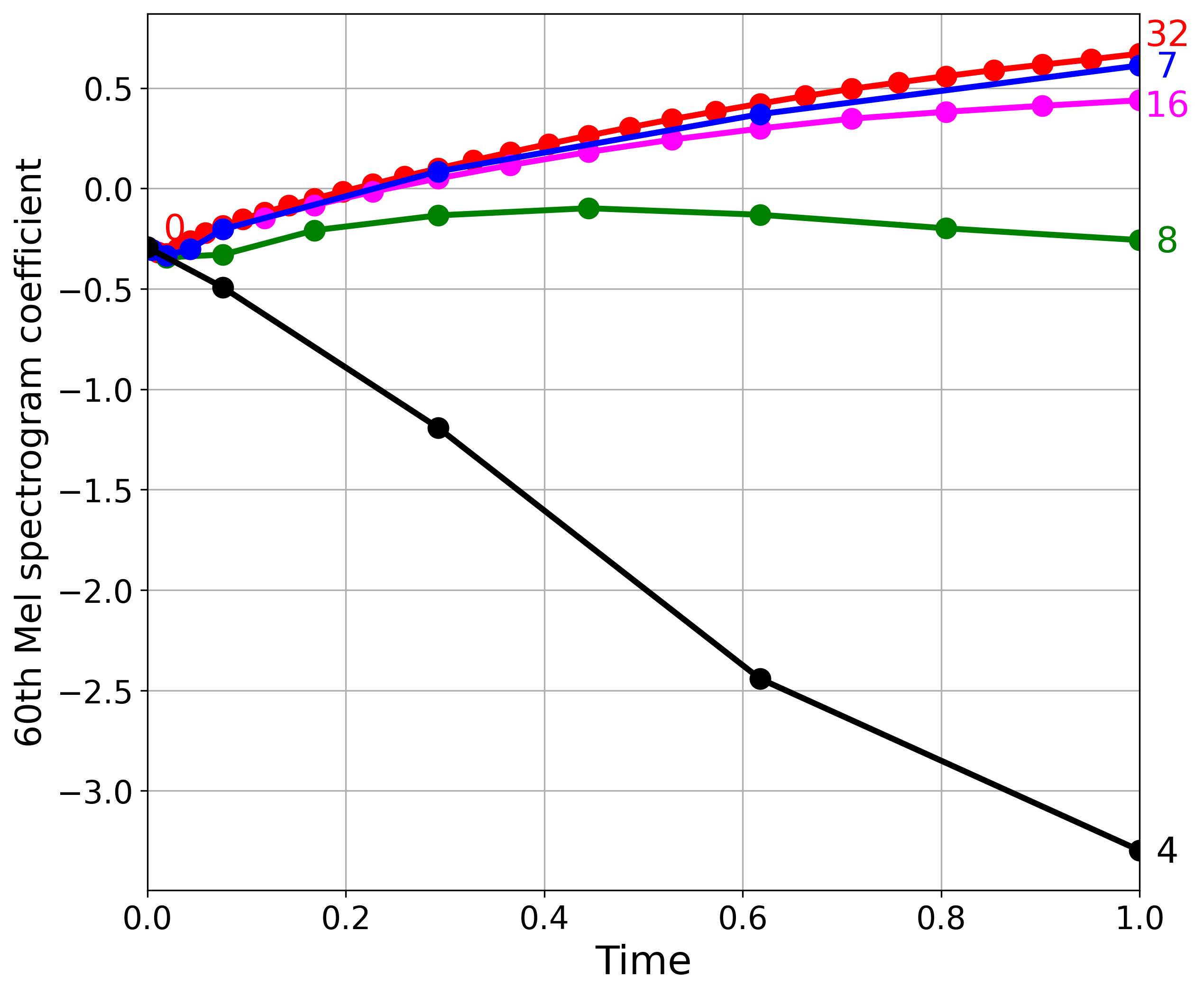} \hfill
  \includegraphics[width=0.32\linewidth]
  {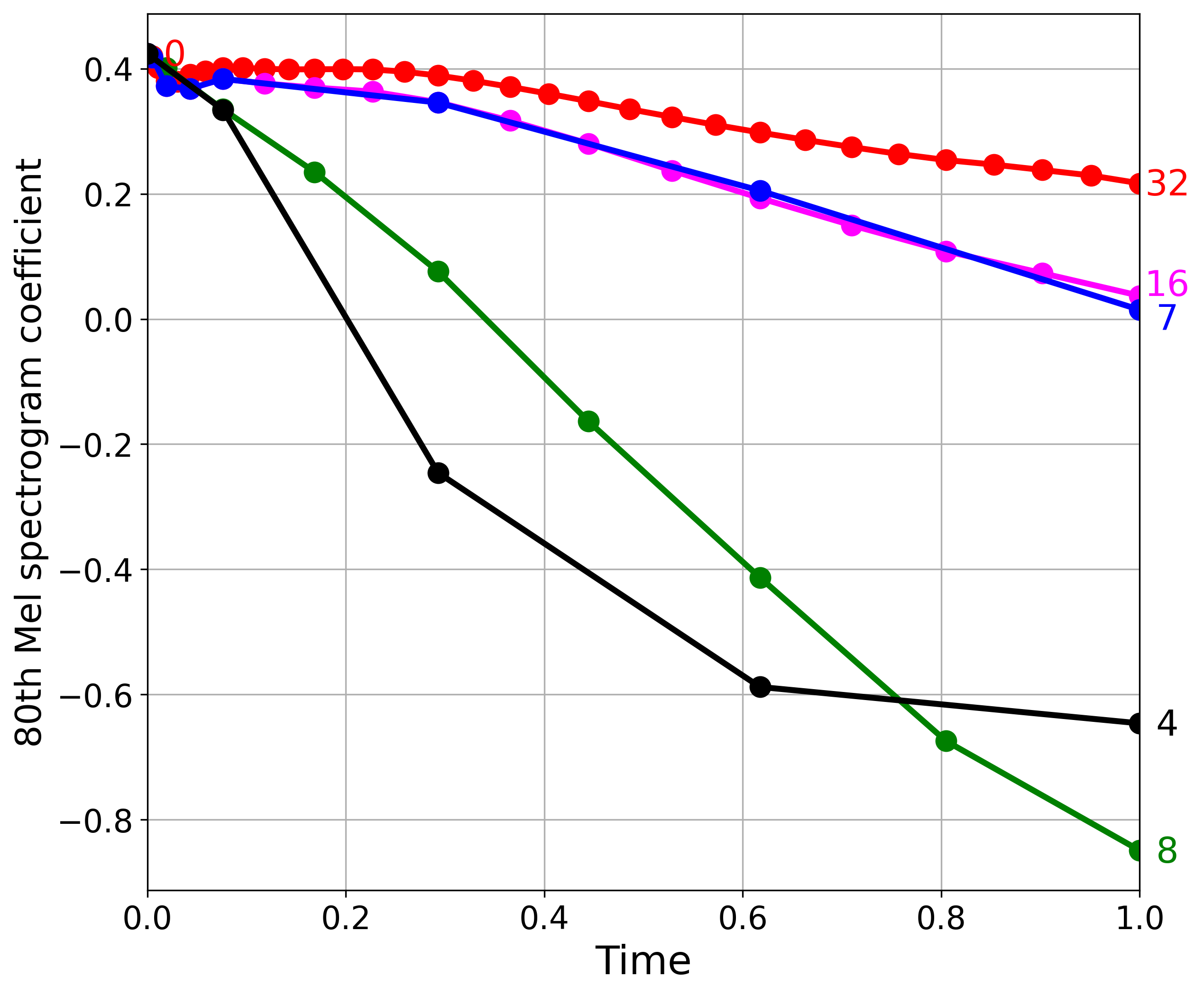}
  \caption {Sampling trajectories for the 20th, 60th, and 80th dimensions of the 100-dimensional Mel spectrogram feature.}
  \label{fig:trajs}
\end{figure*}

\section{Methods}

\subsection{Trajectories analysis of flow matching sampling}
\label{sec:analysis}
Flow-matching-based generative models transform noise into real-world data through multiple neural network inference steps, with the time-step $t$ progressing from 0 to 1. 
Most flow matching models use a uniform time-step schedule for sampling, where $t_k = \frac{k}{\text{NFE}}, k = 0, 1, ..., \text{NFE}$. 
Although this strategy produces high-quality samples, it often requires tens of function evaluations (e.g., 32 NFE in F5-TTS), leading to high computational and time costs.

To optimize this process, we analyze the F5-TTS sampling trajectory where sway sampling is applied.
Figure~\ref{fig:traj_base} shows the Mel spectrogram generation process of a single frame from Gaussian noise. 
By applying PCA to the 100-dimensional Mel features, we reduce the dimensionality for qualitative analysis of the trajectory. Figure~\ref{fig:trajs} plots the trajectories of selected Mel feature dimensions, providing insight into the generation dynamics.

Using the 32-NFE trajectory as an example, two key observations emerge:

\noindent \textbf{Complex initial phase}: The early trajectory exhibits significant curvature, likely due to the model operating on near-noise inputs, leading to uncertainty in the flow direction.

\noindent \textbf{Simplified later phase}: As the trajectory progresses, it becomes nearly linear, indicating that the model inputs now contain sufficient information, allowing the flow's direction to be well-defined.

These trajectory characteristics reflect the properties of the underlying vector field in Equation~\ref{eq: ODE}. 
Specifically, linear vector fields (in the later phase) allow for larger step sizes, while nonlinear fields (in the early phase) require smaller steps and more iterations for precision. 

Additionally, the ODE solving process for flow matching inference can be seen as a Markov process, where the precision of later steps depends on the precision of the early steps. This theoretical foundation emphasizes the importance of accurately solving the initial steps to ensure overall inference precision.

\subsection{Empirically pruned step sampling (EPSS)}

To address the inefficiencies of uniform sampling, we propose EPSS, a non-uniform time-step scheduling method that prunes unnecessary steps based on empirical observations. Guided by the trajectory characteristics analyzed in Section~\ref{sec:analysis}, EPSS allocates denser steps with smaller intervals in the initial phase to preserve quality, while pruning steps in the later phase to accelerate inference.

\begin{figure}[t]
\centering
  \includegraphics[width=0.87\linewidth]{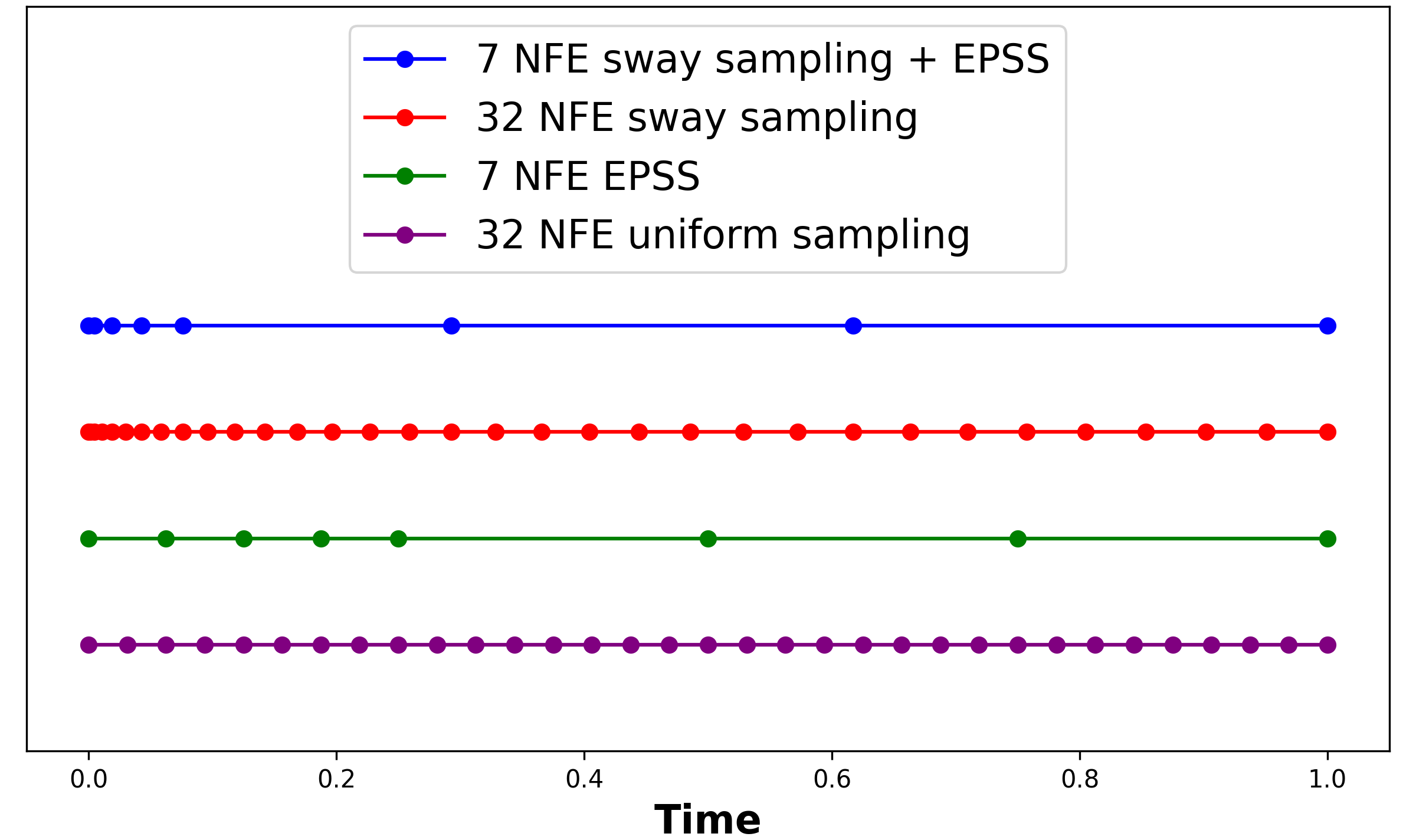}
  \caption {Illustration of EPSS sampling time-step distribution.}
  \label{fig:epss}
\end{figure}

A representative 7-NFE EPSS configuration is shown in Figure~\ref{fig:epss}, with time-steps $\{0, \frac{1}{16}, \frac{1}{8}, \frac{3}{16}, \frac{1}{4}, \frac{1}{2}, \frac{3}{4}, 1\}$.
Compared to the 32-NFE uniform sampling, this configuration reduces computational costs by 78\% while maintaining comparable synthesis quality. As shown in Figure~\ref{fig:traj_base} and Figure~\ref{fig:trajs}, the 7-NFE sway sampling + EPSS trajectory closely matches the 32-NFE baseline, significantly outperforming 8-NFE sway sampling. In Section~\ref{sec:exp}, we validate these results through quantitative metrics and ablation studies.

\section{Experiments}
\label{sec:exp}

\subsection{Setup}

\textbf{Baselines}. We use officially open-sourced F5-TTS \textit{Base}~\cite{chen2024f5} as the baseline, which was trained on the Emilia dataset~\cite{he2024emilia} for 1.2M updates to explore optimal settings for our EPSS method. The reproduced E2 TTS model from \cite{chen2024f5} is included for evaluation and generalization study. 
Besides, we compared the performance of ours with several other diffusion or flow-matching-based TTS systems, including Seed-TTS~\cite{anastassiou2024seed}, MaskGCT~\cite{wang2024maskgct} and CosyVoice 2~\cite{du2024cosyvoice}. We directly report the performance results as reported in their respective papers.

\noindent \textbf{Inference.} We follow the setting of F5-TTS, using the Euler ODE solver, CFG strength $2.0$, sway sampling with coefficient $-1.0$, and pre-trained Vocos~\cite{siuzdak2023vocos} as vocoder. The generation duration is linearly estimated~\cite{chen2024f5}. For easier comparison, we also adopt the Euler ODE solver and sway sampling for E2 TTS while the original setting is the midpoint and uniform sampling.

\subsection{Evaluation}
We adopt Seed-TTS-eval \cite{anastassiou2024seed} and LibriSpeech-PC \textit{test-clean} \cite{meister2023librispeech} as our test set, evaluating performance through four metrics. All measurements are performed on single NVIDIA RTX 3090 GPU to ensure consistent hardware conditions.

\noindent \textbf{Word Error Rate (WER)} measures the intelligibility of synthesized speech by comparing its transcription with the ground-truth text. We employ Whisper-large-V3~\cite{radford2023robust} and Paraformer-zh~\cite{gao2022paraformer} for recognition and compute WER accordingly. 

\noindent \textbf{Speaker Similarity (SIM-o)} reflects the resemblance between the synthesized and ground-truth speech. We use WavLM-TDNN~\cite{chen2022wavlm} to extract speaker embeddings and compute their cosine similarity between them.

\noindent \textbf{UTMOS}~\cite{saeki2022utmos} is an objective metric for evaluating the naturalness of synthesized speech. It employs an open-source MOS prediction model to estimate audio quality without requiring reference recordings or labels. While not an absolute subjective measure, UTMOS provides a practical and efficient way to evaluate the naturalness of synthetic speech.

\noindent \textbf{Real-Time Factor (RTF)} measures the time efficiency of a speech synthesis system by quantifying the time required to generate one second of speech. As there is no standardized benchmark for evaluating RTF, we provide our detailed procedure:
(1) Input a 6-second speech prompt; (2) Use the TTS system to synthesize a 20-second speech segment (excluding the prompt); (3) Repeat this process 100 times; and 
RTF is computed as the ratio of the total inference time cost to the total length of generated speech.
Note that the time cost includes both the Mel spectrogram encoding and the waveform decoding via a vocoder, ensuring end-to-end measurement. RTF exhibits ~10\% variability across runs, even on the same machine.

\subsection{Results}
Table~\ref{tab:main_result_} shows the main results. Compared to the baseline F5-TTS model with 32 NFE, Fast F5-TTS significantly improves inference efficiency while maintaining competitive synthesis quality. 
With fewer NFE, Fast F5-TTS achieves a substantial reduction in RTF from 0.123 to 0.030, showing the effectiveness of EPSS in enhancing speed. 

In Seed-TTS-eval, despite the lower computational cost, Fast F5-TTS maintains comparable or better performance in terms of SIM-o and UTMOS, indicating that speech naturalness and speaker similarity are largely preserved.
Although WER is slightly higher for Fast F5-TTS (1.74 for test-en, 1.75 for test-zh) compared to the baseline (1.70 and 1.58), this modest increase is an acceptable trade-off for the significant gains in inference speed.
Compared to the lower-NFE F5-TTS baselines (16 and 7-NFE), Fast F5-TTS outperforms them in both synthesis quality and efficiency, maintaining competitive or better performance across all metrics while significantly reducing RTF.

The results based on the E2 TTS are consistent with those of F5-TTS, demonstrating that EPSS effectively improves inference efficiency without severely compromising synthesis quality. This consistency across both models highlights the strong generalization ability and broad applicability of our method.

The results on the LibriSpeech-PC \textit{test-clean} further confirm the findings on Seed-TTS-eval, showing similar improvements in inference efficiency and comparable synthesis quality with EPSS, which validates the effectiveness of our method.

\begin{table*} [t]
\caption{Evaluation results on LibriSpeech-PC \textit{test-clean}, Seed-TTS \textit{test-en} and \textit{test-zh}. We report the average results score generated under three different random seeds. The boldface and underline indicate the best and the second-best result respectively. The * denotes reported results from the original papers. $\uparrow$ and $\downarrow$ indicate lower or higher values are better.}
\label{tab:main_result_}
\centering
{\footnotesize
\tabcolsep = 0.7mm
\resizebox{1\linewidth}{!}{
\begin{tabular}{lrccccccccccc}
\toprule
\textbf{Model} & \textbf{\#Param.} & \textbf{NFE} & \textbf{RTF $\downarrow$} & 
\multicolumn{3}{c}{\textbf{LibriSpeech-PC \textit{test-clean}}} & 
\multicolumn{3}{c}{\textbf{Seed-TTS \textit{test-en}}} & 
\multicolumn{3}{c}{\textbf{Seed-TTS \textit{test-zh}}} \\
\cmidrule(lr){5-7} \cmidrule(lr){8-10} \cmidrule(lr){11-13}
 && & & \textbf{WER (\%)$\downarrow$} & \textbf{SIM-o $\uparrow$} & \textbf{UTMOS $\uparrow$} & \textbf{WER (\%)$\downarrow$} & \textbf{SIM-o $\uparrow$} & \textbf{UTMOS $\uparrow$} & \textbf{WER (\%)$\downarrow$} & \textbf{SIM-o $\uparrow$} & \textbf{UTMOS $\uparrow$} \\
\midrule
Ground Truth                         & - & -  & -     & 2.23  & 0.69  & -  & 2.06    & 0.73   & - & 1.26  & 0.76  & - \\
Vocoder Resynthesized                & - & -  & -     & 2.32  & 0.66  & -  & 2.09    & 0.70   & - & 1.27  & 0.72  & - \\
\hdashline[1pt/2pt]\hdashline[0pt/1pt] 
Seed-TTS$_\textit{DiT}$~\cite{anastassiou2024seed}      & - & -  & - & - & - & - & { }{ }\underline{1.73}*  & { }{ }\textbf{0.79}*  & - & { }{ }\textbf{1.18}* & { }{ }\textbf{0.81}* &- \\
MaskGCT~\cite{wang2024maskgct}                      & 1048M & 50 & - & { }{ }2.63* & { }{ }0.69*  & -  & { }{ }2.62*  & { }{ }0.72*  & - & { }{ }2.27* & { }{ }0.77*  & - \\
CosyVoice 2~\cite{du2024cosyvoice}                 & ~500M & 10 & - & { }{ }2.45* & { }{ }\textbf{0.75}*  & - & { }{ }2.57*  & { }{ }0.65*  & - & { }{ }\underline{1.45}* & { }{ }0.75*  & - \\
\hdashline[1pt/2pt]\hdashline[0pt/1pt] 
E2 TTS~\cite{eskimez2024e2}          & 333M & 32 & 0.150 & 2.63  & \underline{0.72} & 3.72 & 1.79  & \underline{0.73} & 3.60 & 1.82  & \underline{0.78} & 2.91 \\
E2 TTS                               & 333M & 16 & 0.078 & 2.78  & 0.71 & 3.71 & 1.88  & \underline{0.73} & 3.61 & 1.93  & \underline{0.78} & 2.93 \\
E2 TTS                               & 333M & 7  & 0.036 & 4.04  & 0.65 & 3.17 & 2.69  & 0.69 & 3.39 & 3.52  & 0.72 & 2.58 \\
\,\, + EPSS                          & 333M & 7  & 0.036 & 2.73  & 0.71 & 3.71 & 1.84  & \underline{0.73} & 3.69 & 1.93  & \underline{0.78} & \underline{2.99}  \\
\hdashline[1pt/2pt]\hdashline[0pt/1pt] 
F5-TTS~\cite{chen2024f5}             & 336M & 32 & 0.123 & \textbf{2.37} & 0.66  & \textbf{3.93} & \textbf{1.70}    & 0.67  & \underline{3.76} & 1.58 & 0.75  & 2.96 \\
F5-TTS                               & 336M & 16 & 0.065 & \underline{2.44} & 0.66  & \underline{3.91} & 1.82    & 0.67  & \textbf{3.78} & 1.75 & 0.75  & 2.97 \\
F5-TTS                               & 336M & 7  & 0.030 & 4.16 & 0.60  & 3.36 & 3.01    & 0.58  & 3.45 & 7.96 & 0.62  & 2.40 \\
\,\, + EPSS ({\bf Fast F5-TTS})      & 336M & 7  & 0.030 & 2.45 & 0.66  & 3.84 & 1.74    & 0.68  & \textbf{3.78} & 1.75 & 0.76  & \textbf{3.00} \\
\bottomrule
\end{tabular}
}
}
\end{table*}

\begin{table} [t]
\caption{\centering
Time-step schedules of EPSS across NFE values. 
$\text{SS}(\cdot)$ is the sway sampling function with coefficient $-1.0$.}
\label{tab:ablation-settings}
\centering
{
\footnotesize
\tabcolsep = 0.7mm
\begin{tabular}{cc}
\toprule
\textbf{NFE} & \textbf{Time-steps} \\
\midrule
32  
& $\text{SS}\left(\frac{1}{32}[\text{0, 1, 2, ..., 32}]\right)$ (F5-TTS Baseline) \\
16  
& $\text{SS}\left(\frac{1}{32}[\text{0, 1, 2, 3, 4, 5, 6, 7, 8, 10, 12, 14, 16, 20, 24, 28, 32}]\right)$ \\
12 
& $\text{SS}\left(\frac{1}{32}[\text{0, { } { } 2, { } { } 4, { } { } 6, { } { } 8, 10, 12, 14, 16, 20, 24, 28, 32}]\right)$ \\
10 
& $\text{SS}\left(\frac{1}{32}[\text{0, { } { } 2, { } { } 4, { } { } 6, { } { } 8, { } { } { } 12, { } { } { } 16, 20, 24, 28, 32}]\right)$ \\
7  
& $\text{SS}\left(\frac{1}{32}[\text{0, { } { } 2, { } { } 4, { } { } 6, { } { } 8, { } { } { } { } { } { } { } { } { } 16, { } { } { } 24, { } { } { } 32}]\right)$ \\
6a 
& $\text{SS}\left(\frac{1}{32}[\text{0, { } { } 2, { } { } 4, { } { } 6, { } { } 8, { } { } { } { } { } { } { } { } { } 16, { } { } { } { } { } { } { } { } { } 32}]\right)$ \\
6b 
& $\text{SS}\left(\frac{1}{32}[\text{0, { } { } 2, { } { } 4, { } { } { } { } { } { } 8, { } { } { } { } { } { } { } { } { } 16, { } { } { } 24, { } { } { } 32}]\right)$ \\
5  
& $\text{SS}\left(\frac{1}{32}[\text{0, { } { } 2, { } { } 4, { } { } 6, { } { } 8, { } { } { } { } { } { } { } { } { } { } { } { } { } { } { } { } { } { } { } { } { } 32}]\right)$ \\
\bottomrule
\end{tabular}
}
\end{table}

\begin{table} [t]
\caption{Ablation results on LibriSpeech-PC \textit{test-clean}.}
\label{tab:ablation-results}
\centering
{
\footnotesize
\tabcolsep = 0.7mm
\begin{tabular}{ccccc}
\toprule
\textbf{NFE} & \textbf{RTF $\downarrow$} & \textbf{WER (\%)$\downarrow$} & \textbf{SIM-o $\uparrow$} & \textbf{UTMOS $\uparrow$} \\
\midrule
32  & 0.123 & 2.37 & 0.66 & 3.93 \\
16  & 0.065 & \textbf{2.29} & \textbf{0.67} & \textbf{3.97} \\
12 & 0.050 & 2.44 & 0.66 & 3.93 \\
10 & 0.042 & 2.44 & 0.66 & 3.92 \\
7  & 0.030 & 2.45 & 0.66 & 3.84 \\
6a & 0.027 & 2.41 & 0.66 & 3.82 \\
6b & 0.027 & 2.71 & 0.64 & 3.79 \\
5  & 0.023 & 2.55 & 0.59 & 2.91 \\
\bottomrule
\end{tabular}
}
\end{table}

\subsection{Ablation Study on Time-step Scheduling}
As shown in Table~\ref{tab:ablation-settings}, we explore how varying time-step schedules impact the performance of Fast F5-TTS, aiming to balance inference efficiency with synthesis quality. 

The ablation results are reported in Table~\ref{tab:ablation-results}. 
As expected, reducing the NFE leads to a significant decrease in RTF, enhancing the inference efficiency. 
At 16 NFE, EPSS outperforms the F5-TTS baseline in all metrics .
At lower NFE, between 7 and 12, EPSS remains a stable performance, with only slight degradation in WER and UTMOS. 
However, performance sharply declines below 6 NFE, indicating limits to reducing time-steps without affecting synthesis quality.

For NFE = 6, we designed two distinct schedules, 6a and 6b, both derived from the 7-NFE setting. Schedule 6a prunes the time-step at $t = \text{SS}(3/4)$, whereas schedule 6b prunes that at $t = \text{SS}(3/16)$. The results illustrate that 6a substantially outperforms 6b in WER. This underscores the strategic importance of selective pruning of time-steps in the later phase, which is in line with our qualitative analysis in Section~\ref{sec:analysis}.

These findings not only validate the strategic pruning approach of EPSS but also highlight its dual capability: maintaining high performance with higher NFE and speeding up the inference at lower NFE levels without significant quality losses, within a certain range. This dual benefit illustrates EPSS's adaptability and effectiveness across different operational settings, making it a robust solution for accelerating flow-matching-based TTS systems.

\subsection{Limitations}

While EPSS has demonstrated significant improvements in inference efficiency for flow-matching-based TTS systems like F5-TTS and E2 TTS, its application and effectiveness across a wider range of flow-matching-based generative models and tasks remain underexplored. Future research should extend the testing of EPSS to various flow-matching-based generative scenarios to confirm its adaptability and efficacy across different contexts. 
Moreover, EPSS is a training-free approach that provides immediate enhancements by reducing unnecessary sampling steps. However, integrating EPSS with training-based techniques, such as distillation, could potentially amplify its benefits. Such integration might not only accelerate the generative process further but also improve the quality of the synthesized contents. 
Exploring these possibilities could lead to broader applications and more robust generative systems.

\section{Conclusion}

In this work, we introduce Fast F5-TTS, a simple and training-free approach to accelerate inference in flow-matching-based TTS models. By analyzing the sampling trajectory of F5-TTS, we propose Empirically Pruned Step Sampling (EPSS), a non-uniform time-step sampling strategy that prunes redundant sampling steps, to reduce the NFE while preserving synthesis quality. Our method, which generates with only 7 NFE, achieves a 4$\times$ speedup over the original F5-TTS without compromising synthesis quality. It also demonstrates strong generalization ability on E2 TTS. The results highlight EPSS as an effective solution for improving the efficiency of flow-matching-based TTS systems with minimal performance degradation. 

\clearpage

\section{Acknowledgements}

This work was supported by the National Natural Science Foundation of China  (No. U23B2018 and No. 62206171), Shanghai Municipal Science and Technology Major Project under Grant 2021SHZDZX0102.
The authors also acknowledge Beijng PARATERA Tech CO.,Ltd. for providing HPC resources that have contributed to the research results reported within this paper. URL: \url{https://paratera.com/}

% \ifinterspeechfinal
%      The Interspeech 2025 organisers
% \else
%      The authors
% \fi
% would like to thank ISCA and the organising committees of past Interspeech conferences for their help and for kindly providing the previous version of this template.

\bibliographystyle{IEEEtran}
\bibliography{mybib}

\end{document}